\documentclass[onecolumn,10pt,nofootinbib,superscriptaddress]{article}
\usepackage{amsmath,amssymb,amsthm,bm,graphicx,mathtools,braket,dcolumn,multirow}
\usepackage{multirow}
\usepackage{enumitem}
\DeclareSymbolFontAlphabet{\amsmathbb}{AMSb}%
\usepackage[unicode=true,
bookmarks=true,bookmarksopen=false,
breaklinks=false,pdfborder={0 0 0},colorlinks=true]
{hyperref}
\usepackage{xcolor}
\definecolor{cblue}{rgb}{0.16, 0.32, 0.75}
\definecolor{cred}{rgb}{0.7, 0.11, 0.11}
\hypersetup{%
	,linkcolor=cred
	,citecolor=cblue
	,urlcolor=black
}
\usepackage[normalem]{ulem}
\usepackage{bm}

\def\<{\langle}
\def\>{\rangle}

\newtheorem{Proposition}{Proposition}
\newtheorem{Corollary}{Corollary}

\newtheorem{Remark}{Remark}

\newtheorem{DEF}{Definition}

\numberwithin{equation}{section}

\date{}

\begin{document}
	
	\title{
{\bf Time inhomogeneous quantum dynamical maps}}

	
	\author{Dariusz Chru\'sci\'nski \\
Institute of Physics, Faculty of Physics, Astronomy and Informatics \\
 Nicolaus Copernicus University\\
  Grudziadzka 5/7, 87-100 Toru\'n, Poland}
	
\maketitle
	
\begin{abstract}
We discuss a wide class of time inhomogeneous quantum evolution which is represented by two-parameter family of completely positive trace-preserving maps. These dynamical maps are constructed as infinite series of jump processes.
It is shown that such dynamical maps satisfy time inhomogeneous memory kernel master equation which provides a generalization of the master equation involving the standard convolution. Time-local (time convolution-less) approach is discussed as well. Finally, the comparative analysis of traditional time homogeneous vs. time inhomogeneous scenario is  provided.

\end{abstract}

\section{Introduction}

The dynamics of an open quantum system \cite{Open-1,Open-2} is usually represented by the dynamical map $\{\Lambda_{t,t_0}\}_{t \geq t_0}$, i.e. a family of completely positive trace-preserving maps $\Lambda_{t,t_0} : \mathcal{B}(\mathcal{H}) \to \mathcal{B}(\mathcal{H})$ \cite{Paulsen,Stormer} ($\mathcal{B}(\mathcal{H})$ stands for the vector space of bounded linear operators acting on the system's Hilbert space $\mathcal{H}$). In this paper we consider only finite dimensional scenario and $\mathcal{B}(\mathcal{H})$ and hence  $\mathcal{B}(\mathcal{H})$ contains all linear operators. The map $\Lambda_{t,t_0}$ transforms any initial system's state  represented by a density operator $\rho_{0}$ at an initial time $t_0$ into a state at the current time $t$, i.e. $\rho_t = \Lambda_{t,t_0}(\rho_{0})$.
Dynamical maps $\{\Lambda_{t,t_0}\}_{t \geq t_0}$ provide the powerful generalization of the standard Schr\"odinger unitary evolution $U_{t,t_0} \rho_{0} U_{t,t_0}^\dagger$, where $U_{t,t_0}$ is a family of unitary operators acting on $\mathcal{H}$.  A dynamical map is usually realized as a reduced evolution \cite{Open-1}

\begin{equation}\label{RED}
  \Lambda_{t,t_0}(\rho_0) = {\rm Tr}_E\left( \mathbb{U}_{t,t_0} \rho_0 \otimes \rho_E \mathbb{U}_{t,t_0} \right) ,
\end{equation}
where $\mathbb{U}_{t,t_0}$ is a unitary operator acting on $\mathcal{H} \otimes \mathcal{H}_E$, $\rho_E$ is a fixed state of the environment (living in $\mathcal{H}_E$), and ${\rm Tr}_E$ denotes a partial trace (over the environmental degrees of freedom). The unitary $\mathbb{U}_{t,t_0}$ is governed by the total (in general time-dependent) `system + environment' Hamiltonian $\mathbb{H}_t$. Now, if $\mathbb{H}_t=\mathbb{H}$ does not depend on time 
the reduced evolution (\ref{RED}) is time homogeneous (or translationally invariant), i.e. $\Lambda_{t,t_0} = \Lambda_{t-t_0}$ (or equivalently $\Lambda_{t+\tau,t_0+\tau} = \Lambda_{t,t_0}$ for any $\tau$). In this case one usually fixes $t_0=0$ and simply considers one-parameter family of maps $\{\Lambda_t\}_{t \geq 0}$.  Such scenario is usually considered by majority of authors. The most prominent example of time homogeneous dynamical maps is the celebrated Markovian semigroup $\Lambda_{t} = e^{\mathcal{L} t}$, where $\mathcal{L}$ denotes the Gorini-Kossakowski-Lindblad-Sudarshan (GKLS) generator \cite{GKS,L} (cf. also the detailed exposition in \cite{Alicki} and  \cite{40-GKLS} for a brief history)

\begin{equation}\label{GKLS}
  \mathcal{L}(\rho) = - i[H,\rho] + \sum_k \gamma_k \left( L_k \rho L_k^\dagger - \frac 12 \{ L_k^\dagger L_k,\rho\} \right) ,
\end{equation}
with the (effective) system's Hamiltonian $H$, noise operators $L_k$, and non-negative transition rates $\gamma_k$. It is well known, however, that semigroup evolution usually requires a series of additional assumptions and approximations like e.g. weak system-environment interaction and separation of natural time scales of the system and environment. Departure from a semigroup scenario calls for more refine approach which attracts a lot of attention in recent years and is intimately connected with quantum non-Markovian memory effects (cf. recent reviews \cite{NM1,NM2,NM3,NM4,five,Lidar,PR,Piilo-I,Piilo-II}). To go beyond dynamical semigroup keeping translational invariance one replaces time independent GKLS generator $\mathcal{L}$ by a memory kernel $\{\mathcal{K}_t\}_{t\geq 0}$ and considers the following dynamical equation

\begin{equation}\label{ME-K}
  \partial_t \Lambda_t = \int_0^t \mathcal{K}_{t-\tau} \circ \Lambda_\tau d\tau = \mathcal{K}_t \ast \Lambda_t \ , \ \ \ \Lambda_{t=0} = {\rm id} ,
\end{equation}
where $A \circ B$ denotes composition of two maps. Equation (\ref{ME-K}) is often referred as Nakajima-Zwanzig master equation \cite{Nakajima,Zwanzig}.
The very structure of the convolution $\mathcal{K}_t \ast \Lambda_t$ does guarantee translational invariance. However, the property of complete positivity of $\Lambda_t$ is notoriously difficult as already observed in \cite{Stenholm,Shabani,Steve}. Time non-local master equation (\ref{ME-K}) were intensively studied by several authors \cite{K1,K2,K3,K4,K5,K6,K7,K8,K8a,K9,K10,K11}.
Since the master equation (\ref{ME-K}) involving the convolution is technically quite involved one usually tries to describe the dynamics in terms of convolution-less time-local approach involving a time dependent generator $\{\mathcal{L}_t\}_{t \geq 0}$  (cf. the recent comparative analysis \cite{Nina-NJP}). Time-local generator $\mathcal{L}_t$ plays a key role in characterizing the property of CP-divisibility which is essential in the analysis of Markovianity. Note, however, that the corresponding propagator $\Lambda_{t,s} = \Lambda_t \Lambda_s^{-1}$ is no longer time homogeneous unless $\mathcal{L}_t$ is time independent.

In this paper we go beyond time homogeneous case and consider the following generalization of (\ref{ME-K})

\begin{equation}\label{ME-KN}
  \partial_t \Lambda_{t,t_0} = \int_{t_0}^t \mathcal{K}_{t,\tau} \circ \Lambda_{\tau,t_0} d\tau  , \ \ \ \Lambda_{t_0,t_0} = {\rm id} ,
\end{equation}
which reduces to (\ref{ME-K}) if $\mathcal{K}_{t,\tau} = \mathcal{K}_{t-\tau}$. Equation (\ref{ME-KN}) may be, therefore, considered as a time inhomogeneous Nakajima-Zwanzig master equation. Such description is essential whenever the `system + environment' Hamiltonian $\mathbb{H}_t$ does depend on time.
Note, that formally if $\mathcal{K}_{t,\tau} = \mathcal{L}_t \delta(\tau)$, then (\ref{ME-KN}) reduces to time-local but inhomogeneous master equation

\begin{equation}\label{}
   \partial_t \Lambda_{t,t_0} = \mathcal{L}_t \circ \Lambda_{t,t_0} , \ \ \ \Lambda_{t_0,t_0} = {\rm id} ,
\end{equation}
and the corresponding solution $\Lambda_{t,t_0}$ is CPTP for all $t\geq t_0$ and arbitrary $t_0 \in \mathbb{R}$ if and only if $\mathcal{L}_t$ is of GKLS form for all $t \in \mathbb{R}$ \cite{Open-1,Open-2,Alicki}. This is just inhomogeneous generalization of semigroup evolution and it is often called an inhomogeneous semigroup \cite{Alicki}. Note, that contrary to homogeneous scenario the time dependent generator  $\mathcal{L}_t$ is defined now for all $t\in \mathbb{R}$ (and not only for $t \geq 0$).

In this paper we propose a particular representation of dynamical maps $\{\Lambda_{t,t_0}\}_{t\geq t_0}$ which by construction satisfy (\ref{ME-KN}). Hence, it may be also considered as a particular construction of a legitimate class of memory kernels $\mathcal{K}_{t,\tau}$ giving rise to CPTP dynamical maps. Clearly, it is not the most general  construction. However, the proposed representation  possesses a natural physical interpretation in terms of quantum jumps.  Time-local (time convolution-less) approach is discussed as well. It turns out that a time dependent generator also depends upon the initial time $t_0$, i.e. one has a two-parameter family of generators $\{\mathcal{L}_{t,t_0}\}_{t\geq t_0}$. Finally, the comparative analysis of traditional time homogeneous vs. time inhomogeneous scenario is  provided.

\section{Time homogeneous evolution}

\subsection{Markovian semigroup}

Consider a Markovian semigroup governed by the time independent master equation

\begin{equation}\label{ME}
  \partial_t \Lambda_{t,t_0} = \mathcal{L} \circ \Lambda_{t,t_0} \ , \ \ \ \Lambda_{t_0,t_0} = {\rm id} ,
\end{equation}
where $\mathcal{L}$ stands for the GKLS generator (\ref{GKLS}), and $t_0$ is an arbitrary initial time. It is clear that since $\mathcal{L}$ does not depend on time the dynamical map depends upon the difference $t-t_0$, i.e. the solution of (\ref{ME}) defines one-parameter semigroup $\Lambda_{t,t_0} = \Lambda_{t-t_0}= e^{(t-t_0)\mathcal{L}}$. Usually, one assumes $t_0=0$ and simply writes $\Lambda_t$. Observe, that any GKLS generator (\ref{GKLS}) can be represented as follows

\begin{equation}\label{}
  \mathcal{L} = \Phi - \mathcal{Z} ,
\end{equation}
where $\Phi,  \mathcal{Z} : \mathcal{B}(\mathcal{H}) \to \mathcal{B}(\mathcal{H})$ are linear maps defined by

\begin{equation}\label{}
  \Phi(\rho) = \sum_k \gamma_k L_k \rho L_k^\dagger , \ \ \ \mathcal{Z}(\rho) = C\rho + \rho C^\dagger ,
\end{equation}
with $C = iH + \frac 12 \sum_k L_k^\dagger L_k$.

\begin{Proposition} \label{PRO-1} The solution of Eq.  (\ref{ME}) can be represented via the following series

\begin{equation}\label{I}
  \Lambda_t = \Lambda^{(0)}_t + \Lambda^{(0)}_t \ast \Phi \circ \Lambda^{(0)}_t + \Lambda^{(0)}_t \ast \Phi \circ \Lambda^{(0)}_t \ast  \Phi \circ \Lambda^{(0)}_t + \ldots ,
\end{equation}
where $\Lambda^{(0)}_t = e^{- \mathcal{Z} t}$.
\end{Proposition}
Proof: let us introduce a {\em perturbation parameter} $\lambda$ and a one-parameter family of generators

\begin{equation}\label{}
  \mathcal{L}^{(\lambda)} := \lambda \Phi - \mathcal{Z} ,
\end{equation}
such that $\mathcal{L} = \mathcal{L}^{(\lambda=1)}$. We find a solution to

\begin{equation}\label{ME-l}
  \partial_t \Lambda_{t} = \mathcal{L}^{(\lambda)} \circ \Lambda_{t} , \ \ \ \Lambda_{t=0} = {\rm id} ,
\end{equation}
as a perturbation series

\begin{equation}\label{s}
  \Lambda_{t} = \Lambda^{(0)}_{t} + \lambda \Lambda^{(1)}_{t} + \lambda^2 \Lambda^{(2)}_{t} + \ldots .
\end{equation}
Inserting the series (\ref{s}) into (\ref{ME-l}) one finds
 the following infinite hierarchy of equations

\begin{eqnarray}
    \partial_t \Lambda^{(0)}_t &=& - Z \circ \Lambda^{(0)}_t ,  \nonumber \\
     \partial_t \Lambda^{(1)}_t &=& - Z \circ \Lambda^{(1)}_t + \Phi \circ \Lambda^{(0)}_t  , \nonumber \\
    &\vdots & \nonumber \label{H1} \\
     \partial_t \Lambda^{(\ell)}_t &=& - Z \circ \Lambda^{(\ell)}_t + \Phi \circ \Lambda^{(\ell -1)}_t ,  \\
     & \vdots & \nonumber
\end{eqnarray}
with initial conditions

\begin{equation}\label{INI}
 \Lambda^{(0)}_{t=0} = {\rm id} \ , \ \ \Lambda^{(\ell)}_{t=0} = 0 \ , \ (\ell \geq 1) .
\end{equation}
It is clear that  $\Lambda^{(0)}_t = e^{- \mathcal{Z}t}$, and

\begin{equation}\label{}
 \Lambda^{(\ell+1)}_{t} = \Lambda^{(0)}_{t} \ast \Phi \circ \Lambda^{(\ell)}_{t} = \Lambda^{(0)}_{t} \ast \underbrace{\Phi \circ \Lambda^{(0)}_{t} \ast \ldots \ast \Phi \circ \Lambda^{(0)}_{t}}_{\ell\ \mbox{terms}} .
\end{equation}
Finally, fixing $\lambda=1$ the series (\ref{s}) reduces to (\ref{I}). \hfill $\Box$

Note, that (\ref{I}) is indeed time homogeneous. One finds

\begin{equation}\label{}
  \Lambda_{t-t_0} = \Lambda^{(0)}_{t- t_0} + \Lambda^{(0)}_{t-t_0} \ast \Phi \circ \Lambda^{(0)}_{t-t_0} + \Lambda^{(0)}_t \ast \Phi \circ \Lambda^{(0)}_t \ast  \Phi \circ \Lambda^{(0)}_t + \ldots ,
\end{equation}
and

\begin{equation}\label{}
  A_{t-t_0} \ast B_{t - t_0} := \int_{t_0}^t A_{t-\tau} \circ B_{\tau-t_0}d \tau = \int_{0}^{t-t_0} A_{t-\tau} \circ B_{\tau}d \tau  ,
\end{equation}
does depend upon `$t-t_0$'. A series (\ref{I}) is an alternative representation for the conventional exponential representation

\begin{eqnarray}\label{exp}
  \Lambda_t = {\rm id} +  \mathcal{L} t  + \frac{t^2}{2} \mathcal{L}^2 + \frac{t^3}{3!} \mathcal{L}^3 + \ldots 
  = {\rm id} + t(\Phi-\mathcal{Z}) + \frac{t^2}{2} (\Phi-\mathcal{Z})^2 + \frac{t^3}{3!} (\Phi-\mathcal{Z})^3 + \ldots .
\end{eqnarray}
Note, that contrary to (\ref{exp}) each term in (\ref{I}) is completely positive and has a clear physical interpretation: an $\ell$th term reads

\begin{equation}\label{}
 \Lambda^{(0)}_{t} \ast \underbrace{\Phi \circ \Lambda^{(0)}_{t} \ast \ldots \ast \Phi \circ \Lambda^{(0)}_{t}}_{\ell\ \mbox{terms}} =
  \int_0^t dt_\ell \, \Lambda^{(0)}_{t-t_\ell} \circ \Phi  \circ \int_0^{t_{\ell}}  dt_{\ell-1}\,\Lambda^{(0)}_{t_{\ell} -t_{\ell-1}} \circ \Phi \ldots \circ \Phi \circ \int_0^{t_2}  dt_1\,  \Lambda^{(0)}_{t_2-t_1} \circ \Phi \circ \Lambda^{(0)}_{t_1} ,
\end{equation}
and it can be interpreted as follows: there are $\ell$ quantum jumps up to time `$t$'  at  $\{t_1 \leq t_2 \leq \ldots \leq t_\ell\}$ represented by a completely positive map $\Phi$. Between jumps the system evolves according to (unperturbed) completely positive maps $ \Lambda^{(0)}_{t_2-t_1},\Lambda^{(0)}_{t_3-t_2}, \ldots , \Lambda^{(0)}_{t_\ell -t_{\ell-1}}$. The series (\ref{I}) represents all possible scenario of $\ell$ jumps for $\ell=0,1,2,\ldots$. By construction, the resulting completely positive map $\Lambda_t$ is also trace-preserving. One often calls (\ref{I}) a {\em quantum jump} representation of a dynamical map \cite{Zoller,Plenio,Car}. Note, however, that truncating (\ref{I}) at any finite $\ell$ violates trace-preservation since processes with more than $\ell$ jumps are not included. The standard exponential representation (\ref{exp}) does not have any clear interpretation. Each separate term $t^k \mathcal{L}^k$ does annihilate the trace but is not completely positive. Only the infinite sum of such terms gives rise to completely positive (and trace-preserving) map.

\begin{Corollary} \label{COR-1} Introducing two completely positive maps $Q_t := \Phi \circ   \Lambda^{(0)}_{t}$ and $\,\mathcal{P}_t :=  \Lambda^{(0)}_{t} \circ \Phi$ a series (\ref{I}) can be rewritten as follows
\begin{eqnarray}\label{QP}
  \Lambda_t  
  &=& \Lambda^{(0)}_{t} + \Lambda^{(0)}_{t} \ast \Big( Q_t +  Q_t \ast Q_t + Q_t \ast Q_t \ast Q_t + \ldots \Big)  \nonumber \\
  &=& \Lambda^{(0)}_{t} + \Big( \mathcal{P}_t +  \mathcal{P}_t \ast \mathcal{P}_t + \mathcal{P}_t \ast \mathcal{P}_t \ast \mathcal{P}_t + \ldots \Big) \ast \Lambda^{(0)}_{t} .
\end{eqnarray}
\end{Corollary}



To summarise: the Markovian semigroup represented in (\ref{I}) is constructed out of the {\em unperturbed} completely positive and trace non-increasing map $\Lambda^{(0)}_t= e^{- \mathcal{Z} t}$ and the jump operator represented by a completely positive map $\Phi$. These two objects are constrained to satisfy ${\rm Tr}\mathcal{L}(\rho) = 0$, where $\mathcal{L} = \Phi - \mathcal{Z}$ defines a GKLS generator.

\subsection{Beyond a semigroup}

How to generalize (\ref{I}) beyond a semigroup such that time homogeneity is preserved? Suppose that $\Lambda^{(0)}_t$ is an arbitrary completely positive and trace non-increasing map satisfying $\Lambda^{(0)}_{t=0} = {\rm id}$. Let $\{\mathcal{Z}_t\}_{t\geq 0}$ be a family of maps such that

\begin{equation}\label{}
  \partial_t \Lambda^{(0)}_t = - \mathcal{Z}_t \ast \Lambda^{(0)}_t ,
\end{equation}
that is, $\mathcal{Z}_t$ is a time non-nonlocal generator of $\Lambda^{(0)}_t$. Note, that $\Lambda^{(0)}_t$ defines a semigroup if and only if $\mathcal{Z}_t = \delta(t) \mathcal{Z}$. Consider a family of jump operators represented by completely positive maps $\{\Phi_t\}_{t \geq 0}$. Define now the following generalization of (\ref{I})

\begin{equation}\label{Ia}
  \Lambda_t = \Lambda^{(0)}_t + \Lambda^{(0)}_t \ast \Phi_t \ast \Lambda^{(0)}_t + \Lambda^{(0)}_t \ast \Phi_t \ast \Lambda^{(0)}_t \ast  \Phi_t \ast \Lambda^{(0)}_t + \ldots ,
\end{equation}
that is, one replaces $\Phi \circ \Lambda^{(0)}_t$ by the convolution $\Phi_t \ast \Lambda^{(0)}_t$. By construction (\ref{Ia}) represents a completely positive map being an infinite sum of  completely positive maps

\begin{equation}\label{Lambda-l}
  \Lambda^{(\ell)}_t = \Lambda^{(0)}_{t} \ast \underbrace{\Phi_t \ast \Lambda^{(0)}_{t} \ast \ldots \ast \Phi_t \ast \Lambda^{(0)}_{t}}_{\ell\ \mbox{terms}} ,  \ \ \ \ell = 1,2,\ldots .
\end{equation}
Also a similar quantum jump interpretation still remains true. One finds

\begin{equation}\label{}
  \Lambda^{(\ell)}_t = \int_0^t dt_\ell \Lambda^{(0)}_{t-t-\ell} \circ \ldots \circ \int_0^{t_3} dt_2\, \Phi_{t_3-t_2} \circ \int_0^{t_2} dt_1\, \Lambda^{(0)}_{t_2-t_1} \circ \int_0^{t_1} d\tau\, \Phi_{t_1-\tau} \circ \Lambda^{(0)}_\tau .
\end{equation}
Between jumps the system evolves according to (unperturbed) completely positive maps $ \Lambda^{(0)}_{t_2-t_1},\Lambda^{(0)}_{t_3-t_2}, \ldots , \Lambda^{(0)}_{t_\ell -t_{\ell-1}}$ which are no longer semigroups. 

\begin{Proposition} \label{PRO-2} The map represented by (\ref{Ia}) satisfies the following memory kernel master equation

\begin{equation}\label{ME-K-1}
  \partial_t \Lambda_t = \mathcal{K}_t \ast \Lambda_t , \ \ \ \Lambda_{t=0} = {\rm id} ,
\end{equation}
where

\begin{equation}\label{}
  \mathcal{K}_t = \Phi_t - \mathcal{Z}_t .
\end{equation}
The map $\Lambda_t$ is trace-preserving if and only if $\mathcal{K}_t$ is trace annihilating.
\end{Proposition}
Proof: the proof goes the same lines as that of Proposition \ref{PRO-1}. Introducing

\begin{equation}\label{}
  \mathcal{K}^{(\lambda)}_t = \lambda \Phi_t - \mathcal{Z}_t ,
\end{equation}
and inserting (\ref{s}) into

\begin{equation}\label{ME-K-2}
  \partial_t \Lambda_t = \mathcal{K}^{(\lambda)}_t \ast \Lambda_t  , \ \ \ \Lambda_{t=0} = {\rm id} ,
\end{equation}
one obtains the following infinite hierarchy of equations

\begin{eqnarray}
    \partial_t \Lambda^{(0)}_t &=& - \mathcal{Z}_t \ast \Lambda^{(0)}_t ,  \nonumber \\
     \partial_t \Lambda^{(1)}_t &=& - \mathcal{Z}_t \ast \Lambda^{(1)}_t + \Phi_t \ast \Lambda^{(0)}_t  , \nonumber \\
    &\vdots & \nonumber \label{H2} \\
     \partial_t \Lambda^{(\ell)}_t &=& - \mathcal{Z}_t \ast \Lambda^{(\ell)}_t + \Phi_t \ast \Lambda^{(\ell -1)}_t ,  \\
     & \vdots & \nonumber
\end{eqnarray}
with initial conditions (\ref{INI}). We show that $\Lambda^{(\ell)}_{t} = \Lambda^{(0)}_t \ast \Lambda^{(\ell-1)}_{t}$ is a solution to (\ref{H2}) which immediately implies (\ref{Lambda-l}). Indeed, one has

\begin{equation}\label{}
   \partial_t \Lambda^{(\ell)}_t =  \partial_t [\Lambda^{(0)}_t \ast \Phi_t \ast \Lambda^{(\ell-1)}_{t}] = \Lambda^{(0)}_{t=0} \circ [\Phi_t \ast \Lambda^{(\ell-1)}_{t}] +  [\partial_t \Lambda^{(0)}_t] \ast \Phi_t \ast \Lambda^{(\ell-1)}_{t} ,
\end{equation}
and hence using $\partial_t \Lambda^{(0)}_t = - \mathcal{Z}_t \ast \Lambda^{(0)}_t$, one obtains

\begin{equation}\label{}
   \partial_t \Lambda^{(\ell)}_t =  \Phi_t \ast \Lambda^{(\ell-1)}_{t} - \mathcal{Z}_t \ast \Lambda^{(0)}_t \ast \Phi_t \ast \Lambda^{(\ell-1)}_{t}  =  \Phi_t \ast \Lambda^{(\ell-1)}_{t} - \mathcal{Z}_t \ast  \Lambda^{(\ell)}_{t} ,
\end{equation}
which proves the claim. \hfill $\Box$

\begin{Remark} Usually on solves the time homogeneous differential equations using the technique of Laplace transform. We provide the alternative proof of Proposition \ref{PRO-2} in the Appendix. Here, we  provided the proof which can be easily generalized to inhomogeneous case where the Laplace transform technique can not be directly applied.
\end{Remark}

\begin{Remark} It is clear that if $\Lambda^{(0)}_t = e^{- \mathcal{Z}t}$ is a semigroup, i.e. $\mathcal{Z}_t = \delta(t) Z$, then $\Phi_t = \delta(t) \Phi$, and hence

\begin{equation}\label{}
  \mathcal{K}_t = \delta(t)(\Phi - \mathcal{Z}) = \delta(t) \, \mathcal{L} .
\end{equation}
\end{Remark}

\begin{Corollary} \label{COR-2} Introducing two completely positive maps $Q_t := \Phi_t \ast   \Lambda^{(0)}_{t}$ and $\,\mathcal{P}_t :=  \Lambda^{(0)}_{t} \ast \Phi_t$ a series (\ref{Ia}) can be rewritten as follows
\begin{eqnarray}\label{QP-a}
  \Lambda_t  
  = \Lambda^{(0)}_{t} + \Lambda^{(0)}_{t} \ast \Big( Q_t +  Q_t \ast Q_t + Q_t \ast Q_t \ast Q_t + \ldots \Big)  ,
\end{eqnarray}
or, equivalently,
\begin{eqnarray}\label{QP-b}
  \Lambda_t  = \Lambda^{(0)}_{t} + \Big( \mathcal{P}_t +  \mathcal{P}_t \ast \mathcal{P}_t + \mathcal{P}_t \ast \mathcal{P}_t \ast \mathcal{P}_t + \ldots \Big) \ast \Lambda^{(0)}_{t} ,
\end{eqnarray}
that is, one has exactly the same representation as in the case of semigroup (\ref{QP}). The only difference is the definition of $Q_t$ and $\mathcal{P}_t$ in terms of $\Phi_t$ and $\Lambda^{(0)}_{t}$. Note, however, that if $\Phi_t = \delta(t)\Phi$, then $\Phi_t \ast \Lambda^{(0)}_{t} = \Phi \circ \Lambda^{(0)}_{t}$, i.e. one recovers the same relation as in Corollary \ref{COR-1}.
\end{Corollary}

\begin{Remark} It should be stressed that even when $\Phi_t$ is not completely positive, but $Q_t = \Phi_t \ast \Lambda^{(0)}_{t}$ is completely positive, then
(\ref{QP-a}) is completely positive. Similarly, when $\,\mathcal{P}_t :=  \Lambda^{(0)}_{t} \ast \Phi_t$ is completely positive, then (\ref{QP-b})
is completely positive. Hence, complete positivity of $\Phi_t$ is sufficient but not necessary for complete positivity of the dynamical map $\Lambda_t$. Note, however, if $\Phi_t$ is not completely positive the intuitive interpretation of the series (\ref{Ia}) in terms of quantum jumps is no longer valid.
\end{Remark}

\section{Time inhomogeneous evolution}

\subsection{Time inhomogeneous semigroup}

Consider now the dynamical map $\{\Lambda_{t,t_0}\}_{t\geq t_0}$ governed by the time dependent master equation

\begin{equation}\label{ME-IN}
  \partial_t \Lambda_{t,t_0} = \mathcal{L}_t \circ \Lambda_{t,t_0} \ , \ \ \ \Lambda_{t_0,t_0} = {\rm id} ,
\end{equation}
where $\mathcal{L}_t$ stands for the time dependent GKLS generator, and $t_0$ is an arbitrary initial time. The corresponding solution has the well known structure

\begin{equation}\label{}
  \Lambda_{t,t_0}  = \mathcal{T} \exp\left( \int_{t_0}^t \mathcal{L}_\tau d \tau \right) ,
\end{equation}
where $\mathcal{T}$ stands for chronological time ordering. The two-parameter family of maps $\{\Lambda_{t,t_0}\}_{t \geq t_0}$ satisfies the following composition law

\begin{equation}\label{}
  \Lambda_{t_3,t_2} \circ \Lambda_{t_2,t_1} = \Lambda_{t_3,t_1} ,
\end{equation}
for any triple $\{t_1, t_2, t_3\}$. This very property is a generalization of the standard (homogeneous) semigroup property

\begin{equation}\label{}
   \Lambda_{t_3-t_2} \circ \Lambda_{t_2-t_1} = \Lambda_{t_3-t_1} ,
\end{equation}
and hence one often calls such maps an inhomogeneous semigroup.

Let us represent the time dependent generator as follows
\begin{equation}\label{}
  \mathcal{L}_t = \Phi_t - \mathcal{Z}_t ,
\end{equation}
where now

\begin{equation}\label{}
  \Phi_t(\rho) = \sum_k \gamma_k(t) L_k(t) \rho L_k^\dagger(t) , \ \ \ \mathcal{Z}_t(\rho) = C(t)\rho + \rho C^\dagger(t) ,
\end{equation}
with $C(t) = iH(t) + \frac 12 \sum_k \gamma_k(t)L_k^\dagger(t) L_k(t)$. To find the corresponding jump representation of $\Lambda_{t,t_0}$ let us introduce the following (inhomogeneous) generalization of the convolution.

\begin{DEF} For any two families of maps $A_{t,t_0}$ and $B_{t,t_0}$

\begin{equation}\label{CON}
  (A \circledast B)_{t,t_0} \equiv A_{t,t_0} \circledast B_{t,t_0} := \int_{t_0}^t A_{t,\tau} \circ B_{\tau,t_0}\, d\tau .
\end{equation}
\end{DEF}
Note, that when $A_{t,t_0}=A_{t-t_0}$ and $B_{t,t_0}=B_{t-t_0}$, then

 \begin{equation}\label{}
  (A \circledast B)_{t,t_0} = \int_{t_0}^t A_{t-\tau} \circ B_{\tau-t_0}\, d\tau = \int_{0}^{t-t_0} A_{t-u} \circ B_{u}\, du = (A \ast B)_{t-t_0} .
\end{equation}

\begin{Proposition} \label{PRO-CON} The convolution (\ref{CON}) is associative

\begin{equation}\label{ASS}
  ([A \circledast B] \circledast C)_{t,t_0} = (A \circledast [B \circledast C])_{t,t_0} ,
\end{equation}
for any thee families $A_{t,t_0},\, B_{t,t_0}$ and $C_{t,t_0}$.
\end{Proposition}
See Appendix for the proof.

\begin{Proposition} \label{PRO-3} The solution to (\ref{ME-IN}) can be represented via the following series

\begin{equation}\label{II}
  \Lambda_{t,t_0} = \Lambda^{(0)}_{t,t_0} + \Lambda^{(0)}_{t,t_0} \circledast (\Phi_t \circ \Lambda^{(0)}_{t,t_0}) + \Lambda^{(0)}_{t,t_0} \circledast (\Phi_t \circ \Lambda^{(0)}_{t,t_0}) \circledast  (\Phi_t \circ \Lambda^{(0)}_{t,t_0}) + \ldots ,
\end{equation}
where $\Lambda^{(0)}_{t,t_0}  = \mathcal{T} \exp\left( - \int_{t_0}^t \mathcal{Z}_\tau d \tau \right)$.
\end{Proposition}
Proof: the proof is a generalization of the proof of Proposition \ref{PRO-1}. Consider the family of generators

\begin{equation}\label{}
  \mathcal{L}^{(\lambda)}_t := \lambda \Phi_t - \mathcal{Z}_t .
\end{equation}
We find a solution to

\begin{equation}\label{ME-2}
  \partial_t \Lambda_{t,t_0} = \mathcal{L}^{(\lambda)}_t \circ \Lambda_{t,t_0} , \ \ \ \Lambda_{t_0,t_0} = {\rm id} ,
\end{equation}
as a perturbation series

\begin{equation}\label{s1}
  \Lambda_{t,t_0} = \Lambda^{(0)}_{t,t_0} + \lambda \Lambda^{(1)}_{t,t_0} + \lambda^2 \Lambda^{(2)}_{t,t_0} + \ldots .
\end{equation}
Inserting the series (\ref{s1}) into (\ref{ME-2}) one finds
the following hierarchy of dynamical equations:

\begin{eqnarray}
    \partial_t \Lambda^{(0)}_{t,t_0} &=& - \mathcal{Z}_t \circ \Lambda^{(0)}_{t,t_0} ,  \nonumber \\
     \partial_t \Lambda^{(1)}_{t,t_0} &=& - \mathcal{Z}_t \circ \Lambda^{(1)}_{t,t_0} + \Phi_t \circ \Lambda^{(0)}_{t,t_0}  , \nonumber \\
    &\vdots & \nonumber \label{H3} \\
     \partial_t \Lambda^{(\ell)}_{t,t_0} &=& - \mathcal{Z}_t \circ \Lambda^{(\ell)}_{t,t_0} + \Phi_t \circ \Lambda^{(\ell -1)}_{t,t_0} ,  \\
     & \vdots & \nonumber
\end{eqnarray}
with initial conditions

\begin{equation}\label{INI-1}
  \Lambda^{(0)}_{t_0,t_0} = {\rm id} , \ \ \ \ \Lambda^{(\ell)}_{t_0,t_0} = 0 \ \ (\ell > 0) .
\end{equation}
Clearly, the above hierarchy provides a generalization of (\ref{H1}) for the inhomogeneous scenario. Now,

\begin{equation}\label{}
  \Lambda^{(0)}_{t,t_0} = \mathcal{T} \exp\left( - \int_{t_0}^t \mathcal{Z}_\tau d\tau \right) ,
\end{equation}
defines an inhomogeneous semigroup which is completely positive (but not trace-preserving). As before it is sufficient to show that

\begin{equation}\label{}
  \Lambda^{(\ell)}_{t,t_0} = \Lambda^{(0)}_{t,t_0} \circledast (\Phi_{t} \circ \Lambda^{(\ell-1)}_{t,t_0}) ,
\end{equation}
solves (\ref{H3}). One finds

\begin{equation}\label{}
  \partial_t \Lambda^{(\ell)}_{t,t_0} = \Lambda^{(0)}_{t,t} \circ \Phi_{t} \circ \Lambda^{(\ell-1)}_{t,t_0} +  [\partial_t\Lambda^{(0)}_{t,t_0}] \circledast (\Phi_{t} \circ \Lambda^{(\ell-1)}_{t,t_0}) .
\end{equation}
Using $\Lambda^{(0)}_{t,t} = {\rm id}$, and $\partial_t\Lambda^{(0)}_{t,t_0} = - \mathcal{Z}_t \circ \Lambda^{(0)}_{t,t_0}$, one gets

\begin{equation}\label{}
  \partial_t \Lambda^{(\ell)}_{t,t_0} =  \Phi_{t} \circ \Lambda^{(\ell-1)}_{t,t_0} -  [\mathcal{Z}_t \circ \Lambda^{(0)}_{t,t_0}] \circledast (\Phi_{t} \circ \Lambda^{(\ell-1)}_{t,t_0})
\end{equation}
and finally, observing that

\begin{equation}\label{}
  [\mathcal{Z}_t \circ \Lambda^{(0)}_{t,t_0}] \circledast (\Phi_{t} \circ \Lambda^{(\ell-1)}_{t,t_0}) = \mathcal{Z}_t \circ \Big[ \Lambda^{(0)}_{t,t_0} \circledast (\Phi_{t} \circ \Lambda^{(\ell-1)}_{t,t_0})\Big] = \mathcal{Z}_t \circ \Lambda^{(\ell)}_{t,t_0} ,
\end{equation}
one completes the proof. \hfill $\Box$

For an alternative proof which does not use properties of the convolution `$\circledast$' cf. Appendix.

\subsection{Beyond an inhomogeneous semigroup}


Suppose now that for any initial time $\Lambda^{(0)}_{t,t_0}$ is an arbitrary completely positive and trace non-increasing map satisfying $\Lambda^{(0)}_{t_0,t_0} = {\rm id}$. Let $\{\mathcal{Z}_{t,t_0}\}_{t\geq t_0}$ be a family of maps such that

\begin{equation}\label{ME-CON}
  \partial_t \Lambda^{(0)}_{t,t_0} = 
  - \mathcal{Z}_{t,t_0} \circledast    \Lambda^{(0)}_{t,t_0} ,
\end{equation}
that is $\{\mathcal{Z}_{t,t_0}\}_{t\geq t_0}$ is a inhomogeneous generalization of $\{\mathcal{Z}_t\}_{t\geq 0}$. Now, $\mathcal{Z}_{t,t_0}$ does not only depends upon the current time `$t$' but also upon the initial time $t_0$. Define the following generalization of (\ref{II})

\begin{equation}\label{IIa}
  \Lambda_{t,t_0} = \Lambda^{(0)}_{t,t_0} + \Lambda^{(0)}_{t,t_0} \circledast \Phi_{t,t_0} \circledast \Lambda^{(0)}_{t,t_0} + \Lambda^{(0)}_{t,t_0} \circledast \Phi_{t,t_0} \circledast \Lambda^{(0)}_{t,t_0} \circledast  \Phi_{t,t_0} \circledast \Lambda^{(0)}_{t,t_0} + \ldots ,
\end{equation}
where $\{\Phi_{t,t_0}\}_{t\geq t_0}$ is a family of completely positive maps which reduces to $\{\Phi_t\}_{t\geq 0}$ in the time homogeneous case.
Hence,  one replaces $\Phi_t \circ \Lambda^{(0)}_{t,t_0}$ by the convolution $\Phi_{t,t_0} \circledast \Lambda^{(0)}_{t,t_0}$. By construction Eq. (\ref{IIa}) represents a completely positive map being an infinite  sum of  completely positive maps

\begin{equation}\label{Lambda-la}
  \Lambda^{(\ell)}_{t,t_0} = \Lambda^{(0)}_{t,t_0} \circledast \underbrace{\Phi_{t,t_0} \circledast \Lambda^{(0)}_{t,t_0} \circledast \ldots \circledast \Phi_{t,t_0} \circledast \Lambda^{(0)}_{t,t_0} }_{\ell\ \mbox{terms}} ,  \ \ \ \ell = 1,2,\ldots .
\end{equation}
Clearly,  quantum jump interpretation still remains true.

\begin{Proposition} \label{PRO-4} The map represented by (\ref{IIa}) satisfies the following memory kernel master equation

\begin{equation}\label{ME-K-3}
  \partial_t \Lambda_{t,t_0} = \mathcal{K}_{t,t_0} \circledast \Lambda_{t,t_0}  , \ \ \ \Lambda_{t_0,t_0} = {\rm id} ,
\end{equation}
where

\begin{equation}\label{}
  \mathcal{K}_{t,t_0} = \Phi_{t,t_0} - \mathcal{Z}_{t,t_0} .
\end{equation}
The map $\Lambda_{t,t_0}$ is trace-preserving if and only if $\mathcal{K}_{t,t_0}$ is trace annihilating.
\end{Proposition}
Proof: the proof goes the same lines as that of Proposition \ref{PRO-2} and \ref{PRO-3}. One easily finds the following hierarchy of equations for maps $\Lambda^{(\ell)}_{t,t_0}$ defining the series (\ref{s1}):

\begin{eqnarray}
    \partial_t \Lambda^{(0)}_{t,t_0} &=& - \mathcal{Z}_{t,t_0} \circledast \Lambda^{(0)}_{t,t_0} ,  \nonumber \\
     \partial_t \Lambda^{(1)}_{t,t_0} &=& - \mathcal{Z}_{t,t_0} \circledast \Lambda^{(1)}_{t,t_0} + \Phi_{t,t_0} \circledast \Lambda^{(0)}_{t,t_0}  , \nonumber \\
    &\vdots & \nonumber \label{H4} \\
     \partial_t \Lambda^{(\ell)}_{t,t_0} &=& - \mathcal{Z}_{t,t_0} \circledast \Lambda^{(\ell)}_{t,t_0} + \Phi_{t,t_0} \circledast \Lambda^{(\ell -1)}_{t,t_0} ,  \\
     & \vdots & \nonumber
\end{eqnarray}
with initial conditions (\ref{INI-1}). Clearly, the above hierarchy provides a generalization of (\ref{H2}) for the inhomogeneous scenario.
It is enough to prove that

\begin{equation}\label{}
  \Lambda^{(\ell)}_{t,t_0} = \Lambda^{(0)}_{t,t_0} \circledast \Phi_{t,t_0} \circledast \Lambda^{(\ell-1)}_{t,t_0}  .
\end{equation}
One  has

\begin{equation}\label{}
  \partial_t \Lambda^{(\ell)}_{t,t_0} =  \Lambda^{(0)}_{t,t} \circ \Phi_{t,t_0} \circledast \Lambda^{(\ell-1)}_{t,t_0}  + [\partial_t \Lambda^{(0)}_{t,t_0}] \circledast \Phi_{t,t_0} \circledast \Lambda^{(\ell-1)}_{t,t_0} .
\end{equation}
Using $\Lambda^{(0)}_{t,t} = {\rm id}$, and $\partial_t\Lambda^{(0)}_{t,t_0} = - \mathcal{Z}_{t,t_0} \circledast \Lambda^{(0)}_{t,t_0}$, one gets

\begin{equation}\label{}
  \partial_t \Lambda^{(\ell)}_{t,t_0} =  \Phi_{t,t_0} \circledast \Lambda^{(\ell-1)}_{t,t_0} -  \mathcal{Z}_{t,t_0} \circledast \Big( \Lambda^{(0)}_{t,t_0} \circledast \Phi_{t,t_0} \circledast \Lambda^{(\ell-1)}_{t,t_0} \Big) ,
\end{equation}
and hence

\begin{equation}\label{}
  \partial_t \Lambda^{(\ell)}_{t,t_0} =  \Phi_{t,t_0} \circledast \Lambda^{(\ell-1)}_{t,t_0} -  \mathcal{Z}_{t,t_0} \circledast \partial_t \Lambda^{(\ell)}_{t,t_0} ,
\end{equation}
which ends  the proof. \hfill $\Box$

\begin{Corollary} \label{COR-2a} Introducing two completely positive maps $Q_{t,t_0} := \Phi_{t,t_0} \circledast   \Lambda^{(0)}_{t,t_0}$ and $\,\mathcal{P}_{t,t_0} :=  \Lambda^{(0)}_{{t,t_0}} \circledast \Phi_{t,t_0}$ a series (\ref{Ia}) can be rewritten as follows
\begin{eqnarray}\label{QP-aa}
  \Lambda_{t,t_0}  
  = \Lambda^{(0)}_{{t,t_0}} + \Lambda^{(0)}_{{t,t_0}} \circledast \Big( Q_{t,t_0} +  Q_{t,t_0} \circledast Q_{t,t_0} + Q_ \circledast Q_{t,t_0} \circledast Q_{t,t_0} + \ldots \Big)  ,
\end{eqnarray}
or, equivalently,
\begin{eqnarray}\label{QP-bb}
  \Lambda_{t,t_0}  = \Lambda^{(0)}_{{t,t_0}} + \Big( \mathcal{P}_{t,t_0} +  \mathcal{P}_{t,t_0} \circledast \mathcal{P}_{t,t_0} + \mathcal{P}_{t,t_0} \circledast \mathcal{P}_{t,t_0} \circledast \mathcal{P}_{t,t_0} + \ldots \Big) \circledast \Lambda^{(0)}_{t,t_0} .
\end{eqnarray}
They reduce to (\ref{QP-a}) and (\ref{QP-b}) in the time homogeneous case.
\end{Corollary}

\section{Time local approach}

Very often describing the evolution of an open system one prefers to use a time-local (or so-called convolutionless (TCL)) approach \cite{Open-1}.
Formally, in the time homogeneous case given a dynamical map $\{\Lambda_t\}_{t\geq 0}$ one defines the corresponding time-local generator $\mathcal{L}_t := [\partial_t \Lambda_t] \circ \Lambda_t^{-1} $ (assuming that $\Lambda_t$ is invertible). This way the map $\Lambda_t$ satisfies

\begin{equation}\label{TCL-1}
  \partial_t \Lambda_t = \mathcal{L}_t \circ \Lambda_ t .
\end{equation}
This procedure might be a bit confusing since (\ref{TCL-1}) coincides with (\ref{ME-IN}) for the inhomogeneous map $\Lambda_{t,t_0}$. To clarify this point let us introduce again an initial time and consider $\Lambda_{t,t_0} = \Lambda_{t-t_0}$. Now, the time-local generator reads

\begin{equation}\label{L10}
  \mathcal{L}_{t-t_0} := [\partial_t \Lambda_{t-t_0}] \circ \Lambda_{t-t_0}^{-1} ,
\end{equation}
that is, the generator does depend upon the initial time \cite{PRL-2010}. It implies that the corresponding propagators

\begin{equation}\label{}
  V_{t,s} := \Lambda_{t-t_0} \circ \Lambda^{-1}_{s-t_0} = \mathcal{T} \exp\left( \int_s^t \mathcal{L}_{\tau - t_0} d\tau \right) =   \mathcal{T} \exp\left( \int_{s-t_0}^{t-t_0} \mathcal{L}_{\tau} d\tau \right) ,
\end{equation}
also does depend upon $t_0$. Clearly, fixing $t_0=0$ this fact is completely hidden. The dependence upon $t_0$ drops out only in the semigroup case when $\mathcal{L}_{t-t_0} = \mathcal{L}$.

Similar analysis may be applied to inhomogeneous scenario as well. Now, instead of convolution (\ref{ME-CON}) one may define a time-local generator

\begin{equation}\label{L20}
  \mathcal{L}_{t,t_0} := [\partial_t \Lambda_{t,t_0}] \circ \Lambda_{t,t_0}^{-1} ,
\end{equation}
such that $\Lambda_{t,t_0}$ satisfies the following inhomogeneous TCL master equation

\begin{equation}\label{TCL-2}
  \partial_t \Lambda_{t,t_0} = \mathcal{L}_{t,t_0} \circ \Lambda_{t,t_0} .
\end{equation}
Again, the corresponding propagator

\begin{equation}\label{}
  V_{t,s} := \Lambda_{t,t_0} \circ \Lambda^{-1}_{s,t_0} = \mathcal{T} \exp\left( \int_s^t \mathcal{L}_{\tau,t_0} d\tau \right)  ,
\end{equation}
also does depend upon $t_0$. Hence, the local composition law

\begin{equation}\label{}
  V_{t,s} \circ V_{s,u} = V_{t,u} ,
\end{equation}
holds only if the above propagators are defined w.r.t. the same initial time. Otherwise, composing the propagators does not have any sense.
Equation (\ref{L20}) reduces to (\ref{ME-IN}) only if $\mathcal{L}_{t,t_0}$ does not depend upon $t_0$. In this case one recovers an inhomogeneous semigroup and $\mathcal{L}_{t,t_0} = \mathcal{L}_{t}$.

\section{Conclusions}

We have constructed a family of time inhomogeneous dynamical maps $\{\Lambda_{t,t_0}\}_{t\geq 0}$  represented by the following infinite series

\begin{equation}\label{}
  \Lambda_{t,t_0} = \Lambda^{(0)}_{t,t_0} + \Lambda^{(1)}_{t,t_0} + \Lambda^{(2)}_{t,t_0} + \ldots ,
\end{equation}
where each single map $\Lambda^{(\ell)}_{t,t_0}$ is completely positive. Moreover, the construction does guarantee that $\Lambda_{t,t_0}$ is trace-preserving. Each map $\Lambda^{(\ell)}_{t,t_0}$  represents a process with $\ell$ quantum jumps occurring in the interval $[t_0,t]$. The `free' evolution (no jumps) corresponds to  $\Lambda^{(0)}_{t,t_0}$. Quantum jumps are represented by a family of completely positive maps $\{\Phi_{t,t_0}\}_{t\geq t_0}$ such that $\Lambda^{(\ell)}_{t,t_0}$ is represented as in the following table

\begin{center}

\begin{tabular}{|l||c|c|}
\hline
 & \mbox{Time homogeneous}  & \mbox{Time inhomogeneous}  \\ \hline\hline
 & & \\
\mbox{General map}  & $\ \ \  \Lambda^{(\ell)}_{t-t_0} = \Lambda^{(0)}_{t-t_0} \ast \Phi_{t-t_0} \ast \Lambda^{(\ell-1)}_{t-t_0}  \ \ $ & $\ \ \ \Lambda^{(\ell)}_{t,t_0} = \Lambda^{(0)}_{t,t_0} \circledast \Phi_{t,t_0} \circledast \Lambda^{(\ell-1)}_{t,t_0} \ \ \ $ \\
 & & \\ \hline
  & & \\
\mbox{Markovian semigroup}  $\ $ & $\Lambda^{(\ell)}_{t-t_0} = \Lambda^{(0)}_{t-t_0} \ast (\Phi \circ \Lambda^{(\ell-1)}_{t-t_0})$ & $\Lambda^{(\ell)}_{t,t_0} = \Lambda^{(0)}_{t,t_0} \circledast (\Phi_{t} \circ \Lambda^{(\ell-1)}_{t,t_0}) $ \\
 & & \\ \hline
\end{tabular}

\end{center}
In the time-homogeneous case the above representation simplifies to

\begin{equation}\label{}
  \Lambda_{t-t_0} = \Lambda^{(0)}_{t-t_0} + \Lambda^{(1)}_{t-t_0} + \Lambda^{(2)}_{t-t_0} + \ldots ,
\end{equation}
with a similar interpretation. The dynamical map $\Lambda_{t,t_0}$ satisfies the corresponding Nakajima-Zwanzig memory kernel master equation or equivalently time-local (TCL) master equation displayed in the table

\begin{center}

\begin{tabular}{|l|c|c|}
\hline
 & \mbox{Time homogeneous}  & \mbox{Time inhomogeneous}  \\ \hline \hline
 & & \\
Memory kernel ME  & $\ \ \ \partial_t \Lambda_{t-t_0} = \mathcal{K}_{t-t_0} \ast \Lambda_{t-t_0}\ \ \ $ & $\ \ \ \partial_t \Lambda_{t,t_0} = \mathcal{K}_{t,t_0} \circledast \Lambda_{t,t_0}\ \ \ $ \\
 & & \\ \hline
  & & \\
Markovian semigroup $\ $ & $\mathcal{K}_{t-t_0} = \delta(t-t_0)\, \mathcal{L}$ & $\mathcal{K}_{t,\tau} = \delta(t-\tau)\, \mathcal{L}_t$  \\
 & & \\ \hline
  & & \\
TCL ME & $\partial_t \Lambda_{t-t_0} = \mathcal{L}_{t-t_0} \circ \Lambda_{t-t_0}$ & $\partial_t \Lambda_{t,t_0} = \mathcal{L}_{t,t_0} \circ \Lambda_{t,t_0}$ \\
 & & \\\hline
  & & \\
TCL generator & $\mathcal{L}_{t-t_0} = [\partial_t \Lambda_{t-t_0}] \circ \Lambda_{t-t_0}^{-1}$ & $\mathcal{L}_{t,t_0} = [\partial_t \Lambda_{t,t_0}] \circ \Lambda_{t,q
t_0}^{-1}$ \\
 & & \\\hline
 & & \\
\mbox{New ME} & $\ \ \partial_t \Lambda_{t-t_0} = \mathbb{K}_{t-t_0} \ast \Lambda_{t-t_0} + \partial_t \Lambda_{t-t_0}^{(0)} \ \ $ & $ \ \ \partial_t \Lambda_{t,t_0} = \mathbb{K}_{t,t_0} \circledast \Lambda_{t,t_0} + \partial_t \Lambda_{t,t_0}^{(0)} \ \ $\\
 & & \\\hline

\end{tabular}

\end{center}
Interestingly, apart from Nakajima-Zwanzing memory kernel master equation the map $\Lambda_{t,t_0}$ satisfies the following dynamical equation

\begin{equation}\label{new}
  \partial_t \Lambda_{t,t_0} = \mathbb{K}_{t,t_0} \circledast \Lambda_{t,t_0} + \partial_t \Lambda_{t,t_0}^{(0)} ,
\end{equation}
where the new kernel $\mathbb{K}_{t,t_0}$ is defined by

\begin{equation}\label{}
  \mathbb{K}_{t,t_0} = \partial_t \mathcal{P}_{t,t_0} = \partial_t[ \Phi_{t,t_0} \circledast \Lambda_{t,t_0}^{(0)}] ,
\end{equation}
that is, it is constructed in terms of the `free' evolution represented by  $\Lambda_{t,t_0}^{(0)}$ and the jump operators $\Phi_{t,t_0}$ (the details of the derivation are presented in the Appendix).

This is very general class of legitimate quantum evolutions and corresponding dynamical equations. It would be interesting to apply the above scheme to discuss time inhomogeneous semi-Markov processes \cite{K6,K7,K10,K13}  and collision models (cf. \cite{CM} for the recent review).

\section*{Acknowledgements}
The work was supported by the Polish National Science Centre project No. 2018/30/A/ST2/00837. I thank Stefano Marcantoni for his remark.  

\appendix

\section{Proof of Proposition \ref{PRO-2}}

The simplest way to solve (\ref{H2}) is to pass to the Laplace Transform (LT) domain:

\begin{equation}\label{}
  \widetilde{F}_s := \int_0^\infty e^{-t s} F_t dt .
\end{equation}
Taking LT of (\ref{H2}) one obtains

\begin{equation}\label{}
  s \widetilde{\Lambda}^{(\ell)}_s - {\rm id} = - \widetilde{\mathcal{Z}}_s \circ  \widetilde{\Lambda}^{(\ell)}_s + \widetilde{\Phi}_s \circ \widetilde{\Lambda}^{(\ell)}_s ,
\end{equation}
and using

\begin{equation}\label{}
  s \widetilde{\Lambda}^{(0)}_s - {\rm id} = - \widetilde{\mathcal{Z}}_s \circ  \widetilde{\Lambda}^{(0)}_s ,
\end{equation}
one finds

\begin{equation}\label{}
  s \widetilde{\Lambda}^{(\ell)}_s  = (s+  \widetilde{Z}_s)^{-1} \circ  \widetilde{\Phi}_s \circ \widetilde{\Lambda}^{(\ell-1)}_s = \widetilde{\Lambda}^{(0)}_s \circ  \widetilde{\Phi}_s \circ \widetilde{\Lambda}^{(\ell-1)}_s,
\end{equation}
and hence going back to the time domain one finally obtains $\Lambda^{(\ell)}_t = \Lambda^{(0)}_t \ast \Phi_t \ast \Lambda^{(\ell-1)}_t$ which implies  (\ref{Lambda-la}).

\section{Proof of Proposition  \ref{PRO-CON}}

One has

\begin{equation}\label{}
   (A \circledast [B \circledast C])_{t,t_0} = \int_{t_0}^t d\tau\, A_{t,\tau} \circ [B \circledast C]_{\tau,t_0} =
    \int_{t_0}^t d\tau\, A_{t,\tau} \circ \int_{t_0}^\tau\, du B_{\tau,u} \circ  C_{u,t_0} ,
\end{equation}
and hence using

\begin{equation}\label{}
  \int_{t_0}^t d\tau \, \int_{t_0}^\tau du \ldots = \int_{t_0}^t du \, \int_{u}^t d\tau  \ldots ,
\end{equation}
one obtains

\begin{equation}\label{}
  (A \circledast [B \circledast C])_{t,t_0} = \int_{t_0}^t du \, \left\{ \int_{u}^t d\tau\, A_{t,\tau} \circ B_{\tau,u} \right\} \circ  C_{u,t_0} =
  \int_{t_0}^t du [A \circledast B]_{t,u} \circ C_{u,t_0} = ([A \circledast B] \circledast C)_{t,t_0} ,
\end{equation}
which ends the proof. \hfill $\Box$

\section{Proof of Proposition \ref{PRO-3}}

One finds

\begin{equation}\label{}
  \Lambda^{(\ell)}_{t,t_0} =  \Lambda^{(0)}_{t,t_0} \circ C^{(\ell)}_{t,t_0} ,
\end{equation}
where $C_{t,t_0}$ satisfies

\begin{equation}\label{}
  \partial_t C^{(\ell)}_{t,t_0} = (\Lambda^{(0)}_{t,t_0})^{-1} \circ \Phi_t \circ \Lambda^{(\ell -1)}_{t,t_0} \ , \ \ \ C^{(\ell)}_{t_0,t_0}=0 ,
\end{equation}
and hence

\begin{equation}\label{}
  C^{(\ell)}_{t,t_0} = \int_{t_0}^t (\Lambda^{(0)}_{\tau,t_0})^{-1} \circ \Phi_\tau \circ \Lambda^{(\ell -1)}_{\tau,t_0} ,
\end{equation}
which eventually gives rise to

\begin{equation}\label{}
  \Lambda^{(\ell)}_{t,t_0} =  \int_{t_0}^t \Lambda^{(0)}_{t,\tau} \circ \Phi_\tau \circ \Lambda^{(\ell -1)}_{\tau,t_0} d\tau = \Lambda^{(0)}_{t,t_0}  \circledast (\Phi_t \circ \Lambda^{(\ell -1)}_{t,t_0}) ,
\end{equation}
where we have used the semigroup property

\begin{equation}\label{}
  \Lambda^{(0)}_{t,t_0}\circ (\Lambda^{(0)}_{\tau,t_0})^{-1} =  \Lambda^{(0)}_{t,t_0}\circ \Lambda^{(0)}_{t_0,\tau} = \Lambda^{(0)}_{t,\tau} .
\end{equation}
Simple iteration leads to

\begin{equation}\label{}
 \Lambda^{(\ell+1)}_{t} = \Lambda^{(0)}_{t} \circledast (\Phi \circ \Lambda^{(\ell)}_{t}) = \Lambda^{(0)}_{t} \circledast \underbrace{(\Phi_t \circ \Lambda^{(0)}_{t}) \circledast \ldots \circledast (\Phi_t \circ \Lambda^{(0)}_{t,t_0})}_{\ell\ \mbox{terms}} .
\end{equation}
which proves (\ref{II}). \hfill $\Box$

\section{Derivation of (\ref{new})}

Defining $\mathcal{P}_{t,t_0} = \Phi_{t,t_0} \circledast \Lambda^{(0)}_{t,t_0}$ one has

\begin{equation}\label{}
  \Lambda_{t,t_0} = \Lambda^{(0)}_{t,t_0} + \Big( \mathcal{P}_{t,t_0} + \mathcal{P}_{t,t_0} \circledast\mathcal{P}_{t,t_0} + \mathcal{P}_{t,t_0} \circledast\mathcal{P}_{t,t_0} \circledast\mathcal{P}_{t,t_0} + \ldots \Big) \circledast \Lambda^{(0)}_{t,t_0} ,
\end{equation}
and hence

\begin{equation}\label{}
  \partial_t\Lambda_{t,t_0} = \partial_t \Lambda^{(0)}_{t,t_0} + \partial_t \Big( \mathcal{P}_{t,t_0} + \mathcal{P}_{t,t_0} \circledast\mathcal{P}_{t,t_0} + \mathcal{P}_{t,t_0} \circledast\mathcal{P}_{t,t_0} \circledast\mathcal{P}_{t,t_0} + \ldots \Big) \circledast \Lambda^{(0)}_{t,t_0} ,
\end{equation}
due to $\mathcal{P}_{t,t}=0$.
%
Now,

\begin{eqnarray}\label{}
  && \partial_t \Big( \mathcal{P}_{t,t_0} + \mathcal{P}_{t,t_0} \circledast\mathcal{P}_{t,t_0} + \mathcal{P}_{t,t_0} \circledast\mathcal{P}_{t,t_0} \circledast\mathcal{P}_{t,t_0} + \ldots \Big) \circledast \Lambda^{(0)}_{t,t_0}  \nonumber \\
  && =  \Big(  \partial_t \mathcal{P}_{t,t_0} + [ \partial_t\mathcal{P}_{t,t_0}] \circledast\mathcal{P}_{t,t_0} + [ \partial_t\mathcal{P}_{t,t_0}] \circledast\mathcal{P}_{t,t_0} \circledast\mathcal{P}_{t,t_0} + \ldots \Big) \circledast \Lambda^{(0)}_{t,t_0} \nonumber\\
  && = [\partial_t \mathcal{P}_{t,t_0}] \circledast \Lambda^{(0)}_{t,t_0}  + [\partial_t \mathcal{P}_{t,t_0}] \circledast \Big( \mathcal{P}_{t,t_0} + \mathcal{P}_{t,t_0} \circledast\mathcal{P}_{t,t_0} + \mathcal{P}_{t,t_0} \circledast\mathcal{P}_{t,t_0} \circledast\mathcal{P}_{t,t_0} + \ldots \Big) \circledast \Lambda^{(0)}_{t,t_0}  \nonumber\\
  && = [\partial_t \mathcal{P}_{t,t_0}] \circledast \Lambda^{(0)}_{t,t_0} ,
\end{eqnarray}
which proves (\ref{new}).


\begin{thebibliography}{1} \bibliographystyle{plain}
		
\bibitem{Open-1} H.-P. Breuer and F. Petruccione,
		\textit{The Theory of Open Quantum Systems} (Oxford Univ. Press, Oxford, 2007).
	
\bibitem{Open-2}  A. Rivas and S. F. Huelga, \textit{Open Quantum Systems. An Introduction} (Springer, Heidelberg, 2011).
		

\bibitem{Paulsen} V. Paulsen, \textit{Completely Bounded Maps and Operator
Algebras} (Cambridge University Press, Cambridge, 2003).



\bibitem{Stormer} E. St{\o}rmer, \textit{Positive Linear Maps of Operator Algebras},
Springer Monographs in Mathematics (Springer, New York,
2013).

\bibitem{GKS} V. Gorini, A. Kossakowski, and E.~C. ~G. Sudarshan, Completely positive dynamical semigroups of N-level systems, J. Math. Phys. {\bf 17}, 821 (1976).
		
\bibitem{L}  G. Lindblad, On the Generators of Quantum Dynamical Semigroups, Comm. Math. Phys. {\bf 48}, 119 (1976).

\bibitem{Alicki} R. Alicki and K. Lendi, \textit{Quantum Dynamical Semigroups and Applications} (Springer, Berlin, 1987).

\bibitem{40-GKLS}   D. Chru\'sci\'nski and S. Pascazio, A Brief History of the GKLS Equation, Open Sys. Inf. Dyn. {\bf 24}, 1740001 (2017).



\bibitem{NM1} \'A. Rivas, S. F. Huelga, and M. B. Plenio, Quantum Non-Markovianity: Characterization, Quantification and Detection,  Rep. Prog.
    Phys. {\bf 77}, 094001 (2014).

\bibitem{NM2} H.-P. Breuer, E.-M. Laine, J. Piilo, and B. Vacchini,
Colloquium: Non-Markovian dynamics in open quantum systems, Rev.
Mod. Phys. {\bf 88}, 021002 (2016).


\bibitem{NM3} I. de Vega and D. Alonso, Dynamics of non-Markovian open quantum systems, Rev. Mod. Phys. {\bf 89}, 015001
(2017).

\bibitem{NM4} L. Li, M. J.W. Hall, and H. M. Wiseman, Concepts of quantum non-Markovianity: a hierarchy, Phys. Rep. {\bf 759}, 1
(2018).

\bibitem{five} V. Reimer, M.R. Wegewijs, K. Nestmann, and M. Pletyukhov, Five approaches to exact open-system dynamics: Complete positivity,
divisibility, and time-dependent observables. J. Chem. Phys. {\bf 151}, 044101 (2019).

\bibitem{Lidar} D.A. Lidar, Lecture Notes on the Theory of Open Quantum Systems, arXiv:1902.00967.

\bibitem{PR} D. Chru\'sci\'nski,  Dynamical maps beyond Markovian regime, arXiv:2209.14902

\bibitem{Piilo-I}  C.F. Li, G.C. Guo, and J. Piilo, Non-Markovian quantum dynamics: What does it mean?,  EPL (Europhysics Letters) {\bf 127},
    50001 (2019).

\bibitem{Piilo-II}  C.F. Li, G.C. Guo, and J. Piilo, Non-Markovian quantum dynamics: What is it good for?,   EPL (Europhysics Letters) {\bf 128},
    30001 (2019).

\bibitem{Nakajima} S. Nakajima, On Quantum Theory of Transport Phenomena, Prog. Theor. Phys. {\bf 20}, 948 (1958);

\bibitem{Zwanzig} R. Zwanzig, Ensemble Method in the Theory of Irreversibility, J. Chem. Phys. {\bf 33}, 1338 (1960).


\bibitem{Stenholm} S. M. Barnet and S. Stenholm,  Hazards of reservoir memory, Phys. Rev. A {\bf 64}, 033808 (2001).

\bibitem{Shabani} A. Shabani and D.A.  Lidar,  Completely positive post-Markovian master equation via a measurement approach. Phys. Rev. A. {\bf 71},
020101 (2005).

\bibitem{Steve} S. Campbell, A. Smirne, L. Mazzola, N. Lo Gullo, B. Vacchini, Th. Busch, and M. Paternostro, Critical assessment of two-qubit post-Markovian master equations
Phys. Rev. A {\bf 85}, 032120  (2012).


\bibitem{K1} A. Budini,  Stochastic representation of a class of non-Markovian completely positive evolutions. Phys. Rev. A. {\bf 69}, 042107 (2004).

\bibitem{K2} A. Budini and P. Grigolini, Non-Markovian non-stationary completely positive open quantum system dynamics, Phys. Rev. A {\bf 80}, 022103 (2009)

\bibitem{K3} J. Wilkie and  Y.M. Wong,  Sufcient conditions for positivity of non-markovian master equations with hermitian generators. J. Phys.
A: Math. Gen. {\bf 42}, 015006 (2009).

\bibitem{K4}  S. Maniscalco and F. Petruccione, Non-Markovian dynamics of a qubit, Phys. Rev. A {\bf 73}, 012111 (2006).

\bibitem{K5} A. Kossakowski and R.  Rebolledo, On the Structure of Generators for Non-Markovian Master Equations, Open Sys. Inf. Dyn. {\bf 16}, 259 (2008).

\bibitem{K6} H.-P. Breuer and B. Vacchini,  Quantum semi-Markov processes. Phys. Rev. Lett. {\bf 101}, 140402 (2008).

\bibitem{K7} H.-P. Breuer and B. Vacchini,  Structure of completely positive quantum master equations with memory kernel. Phys. Rev. E. {\bf 79},
041147 (2009).

\bibitem{K8} B. Vacchini,  Non-markovian master equations from piecewise dynamics, Phys. Rev. A. {\bf 87}, 030101 (2013).

\bibitem{K8a}  D. Chru\'sci\'nski and  A. Kossakowski, From Markovian semigroup to non-Markovian
quantum evolution, EPL {\bf 97},  20005 (2012).


\bibitem{K9} D. Chru\'sci\'nski and A. Kossakowski,  Sufficient conditions for a memory-kernel master equation, Phys. Rev. A. {\bf 94}, 020103(R) (2016).

\bibitem{K10} D. Chru\'sci\'nski and A. Kossakowski, Generalized semi-Markov quantum evolution, Phys. Rev. A {\bf 95}, 042131 (2017).

\bibitem{K11} B. Vacchini,  Generalized master equations leading to completely positive dynamics, Phys. Rev. Lett. {\bf 117}, 230401 (2016).

\bibitem{K12}  S. Lorenzo, F. Ciccarello, and G.M.  Palma,  Class of exact memory-kernel master equations. Phys. Rev. A. {\bf 93}, 052111 (2016)

\bibitem{K13} B. Vacchini, Quantum renewal processes, Sci. Rep. {\bf 10}, 5592 (2020).



\bibitem{Nina-NJP} N. Megier, A.  Smirne, and B.  Vacchini, The interplay between local and non-local master equations:
exact and approximated dynamics, New J. Phys. {\bf 22} 083011 (2020).

		
		
		
		
\bibitem{Zoller}  C.~W. Gardiner and P. Zoller, \textit{Quantum Noice} (Springer-Verlag, Berlin, 1999).
		
\bibitem{Plenio} M.~B. Plenio and P. L. Knight, The quantum-jump approach to dissipative dynamics in quantum optics, Rev. Mod. Phys. {\bf 70}, 101 (1998).
		
\bibitem{Car}  H. J.  Carmichael, \textit{An Open Systems Approach to Quantum Optics}, Berlin Heidelberg New-York: Springer-Verlag (1993).


		
\bibitem{PRL-2010} D. Chru\'sci\'nski and  A. Kossakowski, Non-Markovian Quantum Dynamics: Local versus Nonlocal, Phys. Rev. Lett.  {\bf 104},
    070406 (2010).

\bibitem{CM} F. Ciccarello, S. Lorenzo, V. Giovannetti, G.M Palma, Quantum collision models: open system dynamics from repeated interactions, Phys. Rep. {\bf 954}, 1 (2022).

\end{thebibliography}
\end{document}